\documentclass[citesort,12pt]{article}

\usepackage{graphicx}
\usepackage{subfigure}
\usepackage{float}
\newcommand{\bk}{B \to K^* \gamma}
\renewcommand{\thefootnote}{\fnsymbol{footnote}}
\renewcommand{\theequation}{\arabic{section}.\arabic{equation}}
\newcommand{\bea}{\begin{eqnarray}}
\newcommand{\eea}{\end{eqnarray}}
\newcommand{\ra}{\rightarrow}
\newcommand{\lan}{\langle}
\newcommand{\ran}{\rangle}

\begin{document}


\begin{titlepage}
\begin{flushright}
\begin{tabular}{l}
Preprint LPTHE - Orsay 97/16\\
hep-ph/9707271
\end{tabular}
\end{flushright}
\vskip3.5cm
\begin{center}
  {\Large \bf Notes on $B\to K^* \gamma$ \\}
  \vskip1cm {\large Damir Becirevic}
  \vskip0.2cm
  Laboratoire de Physique Th\'eorique et Hautes Energies\\
Universit\'e de Paris XI, B\^atiment 211, 91405 Orsay Cedex, France\\
  \vskip0.2cm
  \vskip1.8cm
  {\large\bf Abstract:\\[10pt]} \parbox[t]{\textwidth}{
We analyzed two form factors $T_1(q^2)$ and $T_2(q^2)$ which enter the
$B\to K^* \ell^+ \ell^-$ decay rate and are accessible in the lattice
calculations for $B \to K^* \gamma$. The extrapolation to $q^2=0$, necessary
for $B\to K^* \gamma$ is examined. The flat behavior of
$T_2(q^2)$ is shown to be disfavored. From the double constrained unitarity
bounds by
the lattice result at small recoil and
the light cone sum rule one at large recoil, the
pole/double pole fit is preferred with the `pole' masses as free parameters. We
also show a consistency of two approaches in the heavy quark limit. Besides
several byproducts of this analysis, our final (conservative) result for
the form factor that determines $B\to K^* \gamma$ rate is $T(0)=0.15\pm 0.04$.}
  \vskip1cm
{\em }
\end{center}
\end{titlepage}

\renewcommand{\thefootnote}{\arabic{footnote}}
\setcounter{footnote}{0}


\section{Introduction}
\setcounter{equation}{0}

In recent years there has been an enormous effort put into studying the flavor
changing neutral current (FCNC) decays. Unlike the other rare decays, $\bk$ is
more attractive because its branching ratio is proportional to $G_F^2~\alpha$
and
not to $G_F^2~\alpha^2$. Indeed, experimentally it was detected and measured.
Today, the increasing statistics allows better and better accuracy so that the
$10\%$ of
experimental error seems to be achievable quite soon. Last summer, CLEO
\cite{cleo} reported:

\begin{equation}
{\cal B}(\bk)=(4.2\pm 0.8 \pm 0.6)\times 10^{-5}
\end{equation}

\noindent
which makes around $20\%$ of the inclusive branching ratio. Based on $b\to
s\gamma$,
this decay is remarkable from several reasons: it allows the access to the
$V_{ts}V_{tb}$ combination of Cabibbo-Kobayashi-Maskawa matrix elements; at
lowest order in the Standard model (SM) it occurs through loops and since the
contribution of W boson is small, one expects signs of the non-SM physics. That
is the main reason why one tries to calculate this decay as precisely as
possible
in the SM. Once the decay rate is fully determined in SM, a discrepancy in its
comparison with experiment could give the strength of the contribution that
comes from the physics beyond. The inclusive rate
is much more complicated and the results suffer from large theoretical
uncertainties. One relies then on exclusive process. Theoretically, a series of
quark models
\cite{tramp,ov,melikhov} and phenomenological approaches \cite{dominguez}
were employed. At the level of guidance and the phenomenological understanding
they are very useful. But, the correct estimate of the decay rate should be
based
on
QCD.

This decay is described by the effective hamiltonian:
\bea
{\cal H}_{eff}(b\to s\gamma )=-{4 G_F\over {\sqrt 2}} V_{ts}^*V_{tb}
\sum_{i=1}^{8} C_i(\mu) {\cal O}_\mu
\eea
\noindent
which is obtained by systematically integrating out heavy degrees of freedom
from the original theory  while descending on the scales of interest (see for
example \cite{buras,ali}). ${\cal O}_{i}(\mu)$  represent the
complete set of operators and $C_{i}(\mu)$ corresponding Wilson coefficients at
the low hadronic scale $\mu$. The dominant contribution to $\bk$ decay comes
from the ${\cal O}_{7}$ operator (electromagnetic penguin) which is
identified as a short
distance contribution. The other contributions (long distance) were first
questioned in \cite{golowich} and come from the vector dominance possibility
$B\to K^* J/\psi (\psi' \dots)\to K^*\gamma$. In fact, next to leading
calculations
\cite{nlo} show that the contributions of other operators are bigger than it
was anticipated at leading order. The significant long distance part in the
case
under study comes from the operator ${\cal O}_{2}$ \cite{milana}, when a photon
is emitted from the charm line\footnote{For $B\to \rho \gamma$, long distance
effects are far more important. See \cite{khodj}}. This part was recently
calculated by the classical QCD
sum rules \cite{wyler}, and estimated to be a small effect for the branching
ratio\footnote{The mixing with ${\cal O}_2$ operator was also studied for the
inclusive case \cite{voloshin}}. The step which is absolutely necessary is to
precisely evaluate the dominant contribution, and we therefore turn our
attention to ${\cal
O}_{7}$. It is convenient to parametrize the hadronic matrix element for $\bk$
as:
\bea
\label{ffs}
\lan K^*(k,e_{\lambda})\vert \bar s \sigma_{\mu \nu} {(1 + \gamma_5)\over 2}
b\vert B(p)\ran &=& 2 \epsilon_{\mu \alpha \beta \gamma} e_{\lambda}^{* \alpha}
p^{\beta} k^{\gamma} T_1(q^2)  \cr \nonumber
&+& i~\left[ e_{\lambda~\mu}^{*} (m_B^2 - m_{K^*}^2) - (e_{\lambda}\cdot q)(p +
k)_\mu \right] T_2(q^2) \cr \nonumber
&+& i (e_{\lambda}\cdot q) \left[ q^\mu - {q^2 \over m_B^2 - m_{K^*}^2} (p +
k)_\mu \right] T_3(q^2)
\eea
\noindent
where $q=p-k$ and $T_{1,2,3}(q^2)$ are the form factors. $T_{3}(q^2)$ does not
contribute $\bk$ decay rate and will not be considered in what follows. From
$\sigma_{\mu \nu} \gamma_5 = - {i\over 2} \epsilon_{\mu \nu \alpha \beta}
\sigma^{\alpha \beta}$ one obtains $T_1(0)=T_2(0)$. This condition is quite
important for lattice analyses since it represents the important constraint
for the extraction of the form factor at the physical point for the on-shell
photon. The decay rate is then given by:
\bea
\Gamma (\bk)={\alpha G_F^2 \over 8 \pi^4} C_7 (m_b)^2 \vert V_{ts}^*
V_{tb}\vert^2 {(m_{b}^2 + m_{s}^2)\over m_{B}^3} (m_{B}^2 - m_{K^*}^2)^3 \vert
T (0)\vert^2.
\eea
\noindent
where we wrote $T_{1,2}(0) \equiv T(0)$. Now, it is obvious how important it is
to
have the exact value of the form factor. To implement the QCD in the
calculation of $T(0)$, we must have the knowledge of the nonperturbative part
of the
hadronic matrix element. As it is
well known, only the lattice QCD and the QCD sum rules can give some answer on
that part.

\section{Penguins on the lattice}
\setcounter{equation}{0}

A huge effort to calculate this hadronic matrix element was done by lattice
groups. To date, there are four lattice groups \cite{ape,ukqcd,lanl,bhs} that
reported results on the $T(0)$. Recall that lattice calculations involve
numerical simulations: using the euclidean space functional integral
formulation of the theory, corresponding correlators are calculated on finite
lattice ($L,~a$) so that the functional integral becomes multidimensional and
can be handled by Monte-Carlo methods. This in principle allows us to calculate
everything from the first
principles of QCD\footnote{For a good review, see \cite{lect}}. But the
problems
are multifold: {\it e.g.} most of the calculations are performed in the
`quenched'
approximation in which the internal quark loops are left out; the actual
lattices cannot accommodate large quark masses ($a^{-1} \sim 2~GeV (\beta =
6.0),\, 2.8~GeV (\beta =  6.2),\, 3.6~GeV (\beta =  6.4)$). We also have to
inject momenta to $B$ and $K^*$ to
cover the kinematically accessible $q^2$, which requires ever larger lattices.
Problems related to the
treatment of fermions on the lattice are well known. The explicitly
introduced (Wilson) term to the naively
discretized fermionic part of the QCD action, introduces
discretization errors of order ${\cal O}(a)$. Most of the today's simulations
are performed with the Wilson
action and with its improved version {\it i.e.}
Sheikholeslami-Wohlert (SW) action, where the discretization errors are reduced
to ${\cal O}(g_0^2 a)$. However, these errors can be quite disturbing in the
studies of the heavy flavor transitions where the corresponding matrix
elements of bilinear operators pick up the lattice artifacts of order
$\cal{O}({\rm m a})$. Very encouraging is the recent proposal of the ALPHA
collaboration for
the improvement of the SW
action (see for example \cite{luscher}), by which the discretization errors are
reduced to only ${\cal O}(a^2)$. At the same time, this proposal also allows us
to reduce the errors for bilinear operators.
Originally, the improvement was carried out in the
chiral limit, which was very recently extended to the case of non-zero quark
masses in \cite{mrsstt}.
Consequently the errors on form factors values out of zero recoil region are
expected to be considerably reduced.
This program should be used in future simulations\footnote{APE collaboration
already started the lattice study of semileptonic and radiative
heavy to light transitions by implementing the new improvement program. First
results are expected by the end of this year}.

Still the
problem of the small heavy mass on the lattice is ubiquitous and the results
obtained
in such a situation must be extrapolated to the physical $m_B$. In our case,
this can be done with the help of heavy quark symmetry which offers important
scaling laws:
\bea
\label{sclaw}
T_1(q^2_{max})\sim  M^{1/2} ~~,~~  T_2(q^2_{max})\sim {1\over M^{1/2}}
\eea
\noindent
The pole dominance prescription then, for the $q^2$ dependence, gives:
\bea
\label{pole}
T_{1,2}(q^2)={T_{1,2}(0) \over ( 1 - {q^2 \over M_p^2} )^n },
\eea
\noindent
which is compatible with the above scaling laws and $T_1(0)=T_2(0)$, if
$T_2(q^2)$ is flat ($n=0$) and $T_1(q^2)$ being dominated by a pole ($n=1$), or
if $T_2(q^2)$ is pole like and $T_1(q^2)$ double pole ($n=2$), and so on. The
actual precision of the lattice data does not allow for a definite conclusion
on this matter. The reason is clear: Directly 'measured' values of form
factors are obtained for smaller masses and limited range of transfer $\vec q =
\vec p - \vec k$. Even with the bigger statistics, a firm conclusion on the
behavior (\ref{pole}) was not claimed. To some extent, the shapes (\ref{pole})
can be tested to `interpolate' to $q^2=0$ since the direct results are obtained
around this point. Again, to make these tests more reliable, lattice data
should be more precise for several $\vec q \neq 0$ and it is particularly at
this point that
the new program for the improvement is expected to play an important role. The
fitted
values
of the
form factors at $q^2=0$ should be then extrapolated from lattice accessible
{\it heavy mass} to the physical $m_B$. The problem was the lack of such a
scaling law. In the case of $B\to \rho (\pi) \ell \nu$ this scaling law can be
obtained in the heavy quark limit from QCD \cite{chernyak,bb} which turns out
to be
$\sim M^{-3/2}$ for all form factors. The situation
for $\bk$ is (apart from small SU(3) breaking, unimportant for these scaling
laws) the same \cite{abs} and for future analyses, these scaling laws
must be used. As a cross check of the analysis, the form factors are first
extrapolated to
$m_B$ and then to $q^2=0$ by means of (\ref{pole}) and with the help of the
constraint $T_1(0)=T_2(0)$.
Of course that it was a necessary check of consistency when doing the analysis
without the scaling law at
$q^2=0$. But, if
we first extrapolate to the $B$ meson mass with
(\ref{sclaw}), the obtained results are just several points concentrated in the
vicinity of the zero recoil point ($q^{2}_{max}=(m_B-m_{K^*})^2$). These values
are important for studying these form factors in $B\to K^* \ell^+ \ell^-$, but
for
$\bk$ another extrapolation ({\it i.e.} to $q^2=0$) is
needed bringing uncontrollable uncertainties to the final result. Let
us also note
that $T_{1,3} (q^2)$ do not contribute at $q^{2}_{max}$ and, since there is no
momentum injection,
the measure of $T_2(q^{2}_{max})$ is very clean.

\section{Unitarity bounds}
\setcounter{equation}{0}

The behavior of form factors with the values known at several points can in
fact be
constrained by some general principles of the theory. Unitarity bounds on the
form factors were generated for different quantities in heavy to heavy
transitions
\cite{bds1,bds11,caprini}, as well as for the form factors in heavy to light
case \cite{bds2}, by using the
old idea of \cite{oku}. The lattice data in this analysis were included first
in
\cite{lellouch}, and then in \cite{damir}. Here we give a brief outline of the
method applying it to our problem.

\noindent
One starts from the two points correlation function:
\bea
\Pi_{\mu \nu \alpha \beta} (q) &=& i \int d^4 x e^{iqx} \lan 0 \vert {\cal T}
\{ j_{\mu \nu}(x),j^{\dagger}_{\alpha \beta}(0)\} \vert 0 \ran \nonumber \\
&=& P^{-}_{\mu \nu \alpha \beta} \Pi_- (q^2) + P^{+}_{\mu \nu \alpha \beta}
\Pi_+ (q^2)
\eea
\noindent
where the tensor current is $j_{\mu \nu}=(1/2)\bar s \sigma_{\mu \nu} b$, and
projectors:
\bea
P^{-}_{\mu \nu \alpha \beta} &=& {1\over q^2} ( q_{\mu} q_{\beta} g_{\nu
\alpha} + q_{\nu } q_{\alpha} g_{\mu \beta} - q_{\mu} q_{\alpha} g_{\nu \beta}
- q_{\nu}q_{\beta} g_{\mu \alpha}) \nonumber \\
P^{+}_{\mu \nu \alpha \beta} &=& P^{-}_{\mu \nu \alpha \beta}  - g_{\mu \beta}
g_{\nu \alpha} + g_{\mu \alpha} g_{\nu \beta}
\eea
project out positive and negative parity parts which come from $1^+$ and $1^-$
intermediate states respectively. When contracted with $q^\nu q^\beta$, this
general
correlator gives:
\bea
i \int d^4 x e^{iqx} \lan 0 \vert {\cal T} \{
j_{\mu}(x), j^{\dagger}_{\alpha}(0)\} \vert 0 \ran =  ( q_{\mu}q_{\alpha} - q^2
g_{\mu \alpha}) \Pi_- (q^2)
\eea
Similarilly, when we deal with $j_{\mu}^{5}=(1/2)\bar s \sigma_{\mu \nu}q^{\nu}
\gamma_5 b$  we obtain:
\bea
i \int d^4 x e^{iqx} \lan 0 \vert {\cal T} \{
j_{\mu}^{5}(x),j^{5\,\dagger}_{\alpha}(0) \} \vert 0 \ran =  (q_{\mu}q_{\alpha}
- q^2
g_{\mu \alpha}) \Pi_+ (q^2)
\eea
As usual, the functions $\Pi_{\pm}(q^2)$ can be written in the form of
n-subtracted dispersion relations:
\bea
\Pi_{\pm}(q^2) = {(q^2)^n \over \pi} \int {Im \Pi_{\pm}(t) \over t^n (t -
q^2)}dt + \sum_{k=0}^{n-1} c_k (q^2)^k
\eea
Unknown subtraction constants are eliminated by taking derivatives, so that we
finally have:
\bea
\chi_{n}^{\pm} = {1\over n!} {\partial^n \Pi_{\pm}(q^2)\over \partial (q^2)^n}
= {1 \over \pi} \int {Im \Pi_{\pm}(t) \over (t - q^2)^{n+1}}dt\,.
\eea
\noindent
The spectral functions can be obtained from the unitarity relation:
\bea
(q_{\mu}q_{\alpha} - q^2 g_{\mu \alpha}) Im \Pi_{\pm}(t + i\epsilon) = {n_f
\over 2}\sum_{\Gamma} \int d\rho_{\Gamma} (2\pi)^4 \delta (p - p_{\Gamma}) \lan
0
\vert j_{\mu}(0)\vert \Gamma \ran \lan \Gamma \vert j^{\dagger}_{\alpha}(0)
\vert 0 \ran,
\eea
where the sum runs over all possible hadronic states with suitable quantum
numbers and with the phase space integration for each allowed state. In this
sum of positive terms, the contribution from $\vert B K^* \ran$ is:
\bea
Im \Pi_- (t) &\geq& {n_f\over 12\pi t^2} \lambda^{3/2}(t) \theta(t-t_+)\vert
T_1(t)\vert^2 \nonumber \\
Im \Pi_+ (t) &\geq& {n_f\over 24\pi t^2} \lambda^{1/2}(t)
(m_{B}^{2}-m_{K^*}^{2})^2 \theta(t-t_+)\vert T_2(t)\vert^2
\eea
\noindent
where we have used the fact that the matrix element for $BK^*$ production is
described by the same set of form factors but real in the region $t>t_+$
($t_{\pm}=(m_B\pm m_{K^*})^2$ and as usual,  $\lambda (t) = (t - t_+)(t -
t_-)$). Note that we do not take $T_3(t)$ into account as mentioned at the
beginning. Now, we may invoke duality and calculate functions
$\chi^{\pm}_n(Q^2)$ ($Q^2=-q^2$) perturbatively. For this to be valid, we
should be far away from the region where the current can create resonances.
Assuming in this case that $Q^2=0$ is far enough, we obtain to leading order:
\bea
\chi^{-}_{2}(0)&=&{N_c \over 2!\, 16\pi^2 m_b^2}\int_{0}^{1} {x^2 (1-x)^2 (2 u
- 3 (1-x)  - 3 u^2 x)\over (1- x+ x u^2)^2 }dx \cr
& & \cr
\chi^{+}_{3}(0)&=&{N_c \over 3!\, 32\pi^2 m_b^4}\int_{0}^{1} {x^3 (1-x)^3 (3 x
(u^2 - 1) + 3 - 8u) \over (1- x+ x u^2)^3 }dx
\eea
\noindent
where $u=m_s /m_b$ and we take it to be zero or $1/25$. Numerical values are
then:
\bea
(u=0)	\hspace*{1.5cm} \chi_2^- &=& 2.375 \times 10^{-3}/m_b^2 \cr
			\chi_3^+ &=& 2.375 \times 10^{-4}/m_b^4  \cr
& & \cr
(u={1\over  25})\hspace*{1.5cm} \chi_2^- &=& 2.121\times 10^{-3}/m_b^2 \cr
			\chi_3^+ &=& 1.167 \times 10^{-4}/m_b^4
\eea
\noindent
Putting this altogether in dispersion relations we arrive to the following
inequalities:
\bea
\label{ineq}
{1\over 6 \pi^2 \chi_2^-} \int_{t_+}^{\infty} {\lambda^{3/2}(t)\over t^5}\vert
T_1(t)\vert^2 &\leq& 1 \nonumber \\
{(m_{B}^{2}-m_{K^*}^{2})^2 \over 12 \pi^2 \chi_3^+} \int_{t_+}^{\infty}
{\lambda^{1/2}(t)\over t^6}\vert T_2(t)\vert^2 &\leq& 1
\eea
\noindent
To extract the information on the form factors in the physically interresting
region for $\bk$ decay accessible on the lattice, one performs a conformal
mapping:
\bea
\label{preslik}
{1+z\over 1-z} = \sqrt {(m_B+m_{K^*})^2 - t \over 4 N m_B m_{K^*}}
\eea
\noindent
by which the regions $t_{-} < t < t_{+}$ and $0 \leq t \leq  t_{-}$ are mapped
into
the segments of the real axis $z_{min} > z > -1$ and $z_{max} \geq z \geq
z_{min}$ respectively, while two branches of the root
are mapped to upper and lower semicircles of $|z|=1$. $z_{min}$ and $z_{max}$
are given by
\begin{eqnarray}
z_{max}= \frac{m_{B}+m_{\pi}-2\sqrt{Nm_{B}m_{
\pi}}}{m_{B}+m_{\pi}+2\sqrt{Nm_{B}m_{\pi}}}
\hspace*{2cm}
z_{min}=-\left( \frac{\sqrt{N} - 1}{\sqrt{N} +
1}
\right)
\nonumber
\end{eqnarray}
\noindent
where $N$ is a free parameter whose role we explain below. It is worth noting
that for $N=1$, $z_{min}=0$.
Generically rewritten, the transformed inequalities
(\ref{ineq}) are:
\begin{eqnarray}
\label{in5}
\frac{1}{2\pi}\int_{0}^{2\pi} d \theta w_i (\vartheta) \vert T_i (\vartheta)
\vert^2
\leq 1 \hspace*{1.5cm}
{\rm or} \hspace*{1.5cm}
\frac{1}{2\pi i}\int_{\vert z\vert =1}\frac{dz}{z}\vert
\phi_{i}(z)T_{i}(z)\vert^{2}
\leq 1 \hspace*{1.5cm}.
\end{eqnarray}
\noindent
The so called outer functions $w_i (\theta)$ are known and analytic on the
circle,
and the analytic functions $\phi_i(\theta)$ can be found as solutions of the
Dirichlet's problem with $\vert \phi_i (\vartheta)\vert^2 = w_i (\vartheta)$ on
the
circle \cite{duren}. Their explicit forms in our case read:
\begin{eqnarray}
\phi_1 (z) &=& \left( 1 \over {6\pi^2 m_B m_\pi N \chi_{2}^{-}} \right)^{1\over
2} (1
+ z)^2 (1 - z)^{1\over 2} \left[ 1 + z + \frac{m_B + m_\pi}{2\sqrt{Nm_B
m_\pi}}(1-z)\right]^{-5} \left[  1 + z + \frac{1 - z}{\sqrt{N}}\right]^{3\over
2} \cr
& & \cr
\phi_2 (z) &=&  { m_B^2 - m_{K^*}^{2}\over 8 \pi m_{B}^2 m_{K^*}^2
N^2 \sqrt{3 N \chi_{3}^{+}}}
 (1 + z) (1 - z)^{7\over 2} \left[ 1 + z + \frac{m_B +
m_\pi}{2\sqrt{Nm_B m_{K^*}}}(1-z)\right]^{-6} \left[  1 + z + \frac{1 -
z}{\sqrt{N}}\right]^{1\over 2}\cr
& &
\end{eqnarray}
\noindent
The space of analytic functions on the unit disk with
\begin{eqnarray}
\Vert T(\vartheta) \Vert_{L_w^2} = \left\{\frac{1}{2\pi}\int_{0}^{2\pi}
d\vartheta \, w
(\vartheta) \vert T (\vartheta) \vert^2 \right\}^{1\over 2}\  <\  \infty
\end{eqnarray}
\noindent
on the boundary is the well known $H_w^2$ space. To make use of the rich theory
of $H^p$ spaces \cite{duren}, we have to ensure our form factors to be analytic
on the disk {\it i.e.} below the threshold for the $BK^*$ production. The
problem with the poles at $B^*_s$ ($5.42~GeV$ \cite{pdg}) and $B_{s1}$
($5.67~GeV$ \cite{hill}) is solved by multiplying the form factors by Blaschke
functions which simply remove the poles sending them on the circle (moduli of
these functions are equal to one):
\bea
T_i (t)\ra    B(z_{pole_i})\, T_i(t) \,=\, {z - z_{pole_i} \over 1 - z
z_{pole_i}}
\,T_i(t)
\eea
\noindent
Besides, the subthreshold singularities must be taken into account. This cannot
be done in a model independent
way. Some reasonable model for the cut must be employed. Such
analyses were performed in \cite{bds11} demonstrating that the effect of these
singularities is
small and that the final bounds are just few percents weaker. While this is
very important for this sort of analyses for some subtle quantities, like the
slope and the curvature of the Isgur-Wise function in $B\to D^* \ell \nu$ (for
the
latest results in this field, see detailed analysis in \cite{newgbl}), in our
case few percents are really unimportant.
 Now, to generate the bounds we may use different approaches. Here we
adopt one from \cite{caprini,guiasu} which is particularly elegant. The
solution is presented in appendix A. The resulting bounds for the case when the
form factor value is known at one point $z=0$ is:
\bea
{1\over z^2}\left(- \phi_i(0) T_i(0) + \phi_i(z) T_i(z) (1-z^2)\right)^2 +
\phi_i(0)^2 T_i(z)^2 (1-z^2) \leq 1
\eea
When the value is known at two points $z=\{0,z_0\}$ the bounds are
solutions of:
\bea
\label{bds3}
& &{1\over (z-z_0)^2}\left\{ {1\over z_0^2 z^2} \left[ \phi_i(0) T_i(0) (z_0 -
z) + \phi_i(z_0) T_i(z_0) z (1 - z_0 z) (1 - z_0^2) \right. \right. \nonumber
\\  & & \left. \left. - \phi_i(z) T_i(z) z_0 ( 1- z_0 z)  ( 1 - z^2) \right]^2
+  \phi_i(z_0)^2 T_i(z_0)^2 (1 - z_0 z)^2 (1-z_0^2)  \right. \nonumber \\  & &
\left.  + \phi_i(z)^2 T_i(z)^2 (1 - z_0 z)^2 ( 1- z^2) -  2 \phi_i(z_0)
\phi_i(z) T_i(z_0) T_i(z) (1-z_0^2)(1-z^2) (1 - z_0 z) \right\} \leq 1\nonumber
\\ & &
\eea
\noindent
Now when we have the expressions for the bounds, we can apply them to the case
at hand.
To do this, let us remark that the coefficient $N$ in (\ref{preslik}) is free
and for commodity we can always choose it in such a way that the value of the
form factor is known at $z=0$. So the formulae for the bounds are always
applicable by a simple adjustment of $N$. Here we use APE results for
$T_{1,2}(t)$
(technical details can
be found in \cite{ape}). After the extrapolation to $m_B$, the results that we
use are:
\bea
T_2(19.2)= 0.25 \pm 0.02~~(0.26\pm 0.02) \nonumber \\
T_2(16.8)=0.24 \pm 0.04~~(0.27\pm 0.05) \nonumber \\
T_1(16.8)=0.56 \pm0.08~~(0.59\pm 0.13)
\eea
\noindent
The above values are obtained from linear (quadratic) fit. The possibility to
extrapolate the results obtained at nonzero recoil point (but in its vicinity)
by the heavy quark
scaling law was explained in \cite{ukqcd}.

\begin{figure}
\centering
\begin{minipage}[h]{\textwidth}
\centering
\subfigure[]{\includegraphics[width=11cm]{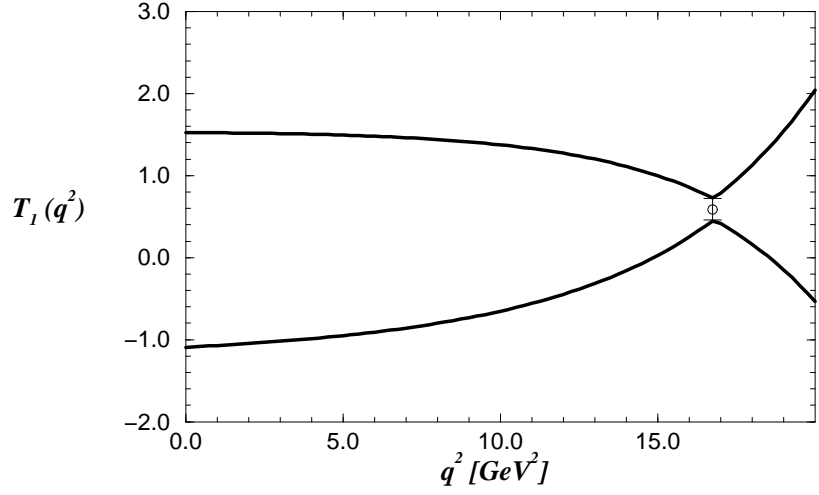}}
\end{minipage}
\centering
\begin{minipage}[h]{\textwidth}
\centering
\subfigure[]{\includegraphics[width=11cm]{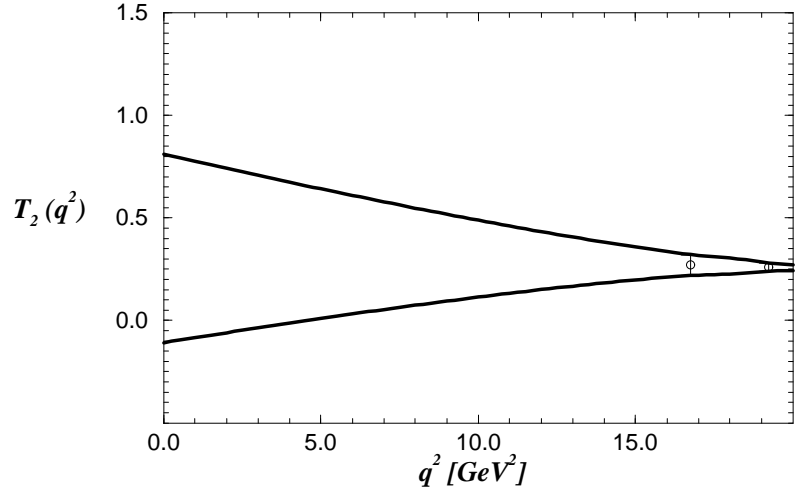}}
\end{minipage}
\caption{\it The lattice constrained unitarity bounds on $T_1(q^2)$ and
$T_2(q^2)$ form factors. The APE lattice point(s) used as constraints are
marked too.}
\end{figure}
\noindent

The typical bounds for the form factors $T_1(t)$ and $T_2(t)$ are depicted in
Fig.1. The bounds generated by using only the lattice results in this fashion,
do not
give satisfactory results, but
they can be more and more useful with the results of forthcoming simulations,
{\it i.e.} with a bigger range of $q^2$ where the form factors are measured
with a reasonable precision. For the coherent and immediate usefulness, the
bounds should be formulated for
lattice accessible masses. Then, they could even serve to interpolate to
$q^2=0$. We plan to return to this question
soon. For the moment, the bounds on the form factors in the large recoil region
are so wide that not even a hint
on the form factors' behavior can be obtained. However, it should be emphasized
that the situation from Fig.1 is expected. The $q^2$-region accessible from
this decay is so large, that one cannot expect a miracle when constraining by
one or two points at one end of this region. Since we want to learn something
more on the functional
dependence of our form factors, we obviously need some information in the large
recoil region. Before turning to
that point, let us give some more remarks.
Apart from useful scaling laws, the symmetry of heavy flavors opened the
possibility to
 link this decay to
the decays generated by $V-A$ current \cite{wise}. Namely, by using the
equation of
motion for the heavy quark in its rest frame ($\gamma^0 b = b$), we have
 $\bar s \sigma_{0 i} b = - i \bar s \gamma_i (1 - \gamma_5) b$. This allows
for several relations among
 different form factors, valid in the small recoil region (the one accessible
on the lattice).
This fact was not used enough in the previous studies. It is also
interesting to note that in the same limit, the authors of Ref.\cite{oliver}
demonstrated that the
axial current couples only to $_{1\over 2}1^+$ and not to $_{3\over 2}1^+$,
which proves our assumption on the pole for $T_2 (t)$ in this section.

\section{Light cone sum rules}
\setcounter{equation}{0}
In seeking to answer the question on the form factor behavior, one must try to
have
some information in the large recoil region. This is the region where $QCD$ sum
rules proved to be successful.
Here we recalculate $T_1 (q^2)$ and $T_2 (q^2)$ paying some more attention to
the
numerics. In what follows we point out the differences with the papers where
these calculations were originally performed.

$T_1 (0)$ was extensively studied in the sum rule approach
\cite{abs,ball, narison, dpr}. In the case of $B\to K^* \ell^+ \ell^-$ these
form
factors were calculated by means of the classical sum rule technique
\cite{colangelo},
as well as in the light cone (LC) approach \cite{aliev}. The authors of Ref.
\cite{bb} calculated the form factors
for $B\to \rho \ell \nu$ decay and investigated the discrepancies between two
approaches, concluding that the LC
sum rules are more reliable in the large recoil region. LC sum rules seem to be
more useful for us also because they
have well defined heavy quark limit \cite{abs}. Brief and (in our
opinion) the nicest review on LC sum rules in heavy to light processes, can be
found in \cite{braun}. In the
interresting region, the recoiling $s$ quark is very energetic and two
competing mechanisms are on the stage: hard and soft. The hard one comes from
the configuration where the constituents are at small transverse distances so
that the hard gluons are exchanged. This leads to the counting rule which in
the heavy quark limit gives $T_i(0) \sim m_H^{-3/2}$. The soft part comes from
the so called end-point configuration, where almost all the hadron momentum is
carried by $s$ quark so that the transverse separations are not constrained any
more and the nonperturbative technique must be used. In \cite{chernyak}, it was
shown that the soft contribution has the same scaling behavior ({\it i.e.} in
our case $T_i(0) \sim
m_H^{-3/2}$).

\noindent
The sum rule can be derived by considering the correlator:
\bea
F_{\mu}(q,k)&=& i \int d^4 x e^{i q x} \lan K^*(k,e_{\lambda})\vert {\cal T}\{
\frac{1}{2} \bar s(x) \sigma_{\mu \nu} q^{\nu} b(x),\bar b(0) i \gamma_5
d(0)\} \vert 0\ran \cr
&=& \epsilon_{\mu\nu\alpha\beta} e^{* \nu}_{\lambda} q^{\alpha}k^{\beta}
F(q^2,(k+q)^2)
\eea
Formally, the hadronic representation
of this correlator can be obtained by inserting the complete set of states.
With
$\lan B(k+q)\vert \bar b i \gamma_5 d \vert 0\ran = f_B m_{B}^2 / m_b$ and
(\ref{ffs}), we can write down the dispersion relation for the amplitude
$F(q^2,(k+q)^2)$:
\bea
\label{dr}
F(q^2,(k+q)^2) = \int_{m_B^2}^{\infty} \frac{\rho (q^2,s)}{s-(k+q)^2}
ds~+~subtraction
\eea
where the spectral density is written as the $B$ ground state contribution plus
a contribution coming from the higher excitations in that channel:
\bea
\rho  (s,q^2) = 2 T_1(q^2) \frac{f_B m_{B}^{2}}{m_b} \delta (s - m_{B}^{2}) +
\rho^{exc}(q^2,s)
\eea
In the same manner, by using the current $(1/2)\bar
s \sigma_{\mu \nu} q^{\nu}\gamma_5 b$, for the second form factor we have:
\bea
\tilde F_{\mu}(q,k)&=&e^{*}_{(\lambda) \mu} \tilde F(q^2,(k+q)^2) ~~{\rm
and}\nonumber\\
\tilde \rho  (s,q^2) &=& i T_2(q^2)(m_{B}^2 - m_{K^*}^2) \frac{f_B
m_{B}^{2}}{m_b}  \delta (s - m_{B}^{2}) + \tilde  \rho^{exc}(q^2,s)
\eea
On the other hand, the amplitude $F(\tilde F)$ obtained from QCD, satisfy the
same dispersion relation (\ref{dr}), starting from $m_{b}^2$. Then, one invokes
duality of the two representations in the region above a threshold $s_0$ (which
is not a priori known), and equalizes both representations to end up with:
\bea
\frac{2 T_1(q^2)}{m_B^2 - (k+q)^2} \frac{f_B m_{B}^{2}}{m_b} &=& {1\over \pi}
\int_{m_b^2}^{s_0} \frac{Im F^{QCD}(q^2,s)}{s-(k+q)^2} ds \nonumber \\
\frac{i T_2(q^2)}{m_B^2 - (k+q)^2} \frac{f_B m_{B}^{2}}{m_b}(m_{B}^2 -
m_{K^*}^2) &=& {1\over \pi} \int_{m_b^2}^{s_0} \frac{Im \tilde
F^{QCD}(q^2,s)}{s-(k+q)^2} ds
\eea
\noindent
The correlator is defined in such a way that only single dispersion relation is
to be used. In the classical approach, both mesons are treated symmetrically in
a double dispersion relation, so that two threshold parameters are in the game.
Here we have only one, and this can be viewed as one of the important
advantages of the
LC approach. QCD part can be
calculated by expanding the ${\cal T}$ product of the currents near the light
cone $x^2=0$ for the large space-like momenta $(k+q)^2 <0$ (when the virtuality
of
the b-quark is large). Thus, one uses the perturbative expansion of the b-quark
propagator in the external field of slowly varying fluctuations inside the
$K^*$ meson. The remaining matrix elements define the light cone wave
functions. To leading order (bare $b$-quark propagator) one deals with the
following wave functions \cite{zhitn,abs}:
\begin{eqnarray}
\langle K^* (k,\lambda) |\bar s(0)\sigma_{\mu\nu}d(x)
|0\rangle & = &
-if_{K^*}^\perp(\mu)(e^{*(\lambda)}_\mu k_\nu -e^{*(\lambda)}_\nu k_\mu)
 \int_0^1 du\, e^{iukx} \phi_\perp(u,\mu),
\label{}\\
\langle K^*(k,\lambda) |\bar s(0)\gamma_\mu d(x)
|0\rangle &=& k_\mu \frac{(e^{*(\lambda)} x)}{(kx)}
f_{K^*} m_{K^*} \int_0^1 du\, e^{iukx} \phi_\parallel(u,\mu)
\nonumber\\
&&\mbox{}+\left( e^{*(\lambda)}_\mu -k_\mu \frac{(e^{*(\lambda)} x)}{(kx)}
\right) f_{K^*} m_{K^*}
\int_0^1 du\, e^{iukx} g_\perp^{(v)}(u,\mu)\,
\label{}\\
\langle K^*(k,\lambda) |\bar s(0)\gamma_\mu\gamma_5 d(x)
|0\rangle &=&
\frac{1}{4} \epsilon_{\mu\nu\alpha\beta} e^{*(\lambda) \nu}
k^\alpha x^\beta  f_{K^*} m_{K^*}
\int_0^1 du\, e^{iukx} g_\perp^{(a)}(u,\mu)\,.
\label{def3}
\end{eqnarray}
\noindent
$u$ is the momentum fraction
carried by the valence $s$ quark in the $K^*$. In the above definitions, the
gauge phase factor between the quark fields is implicit to ensure the gauge
invariance. In the Fock-Schwinger gauge, above definitions are of course,
exact.
All distribution amplitudes are normalized to one. Here to leading order
means to leading twist (defined as a
difference between the dimension and the Lorentz spin of an operator). To
leading twist - 2, the distribution functions are listed in appendix B as well
as numerical parameters taken for the analysis. It is very important to note
that one of distribution functions that we use is
qualitatively different from the corresponding one used in \cite{abs,aliev}.
Namely, in the
recent study \cite{wfns}, the
distribution amplitude for the transversely polarized $\rho$ meson was shown to
be
broader then the original findings of
\cite{zhitn} argued. We take the functions from \cite{wfns} and account for the
SU(3) breaking. In Fig.2, we display the transverse wave
function which differs from the previous studies. Using the above definitions,
for the QCD part we obtain:
\bea
F^{QCD} (q^2,(k+q)^2) &=& -{i\over 2} \int_{0}^{1} \frac{f_{K^*}
m_{K^*}~du}{m_b^2 - \bar u q^2 + u\bar u m_{K^*}^2 - u (k+q)^2}\times
\nonumber\\
& &\left\{ \frac{m_b f_{K^*}^{\perp}}{f_{K^*} m_{K^*}} \phi_{\perp}(u,\mu)\, +
\, \Phi_{\parallel}(u,\mu) + u
g_\perp^{(v)}(u,\mu)\, +\, \frac{1}{4} g_\perp^{(a)}(u,\mu)\right. \nonumber\\
 &+& \left. \, \frac{m_b^2 + q^2 - u^2
m_{K^*}^2}{4 (m_b^2 - \bar u q^2 + u\bar u m_{K^*}^2 - u (k+q)^2)}
g_\perp^{(a)}(u,\mu) \right\}
\eea
\noindent
and
\bea
\tilde F^{QCD} (q^2,(k+q)^2)&=&{1\over 2} \int_{0}^{1} \frac{f_{K^*}
m_{K^*}~du}{m_b^2 - \bar u q^2 + u\bar u m_{K^*}^2 - u (k+q)^2}\times
\nonumber\\
& &\left\{ (q^2 + u k q)\, g_\perp^{(v)}(u,\mu)\, -\, \frac{u \, [m_{K^*}^2 q^2
-
(kq)^2]}{2 (m_b^2 - \bar u q^2 + u\bar u m_{K^*}^2 - u (k+q)^2)}\,
g_\perp^{(a)}(u,\mu)\,  \right. \nonumber \\
 &+&\left. (k q) \Phi_{\parallel}(u,\mu)\, +\, \frac{m_b f_{K^*}^{\perp}
(kq)}{f_{K^*} m_{K^*}}
\phi_{\perp}(u,\mu)\right\}
\eea
\noindent

\begin{figure}
\centering
\includegraphics[width=6.5cm]{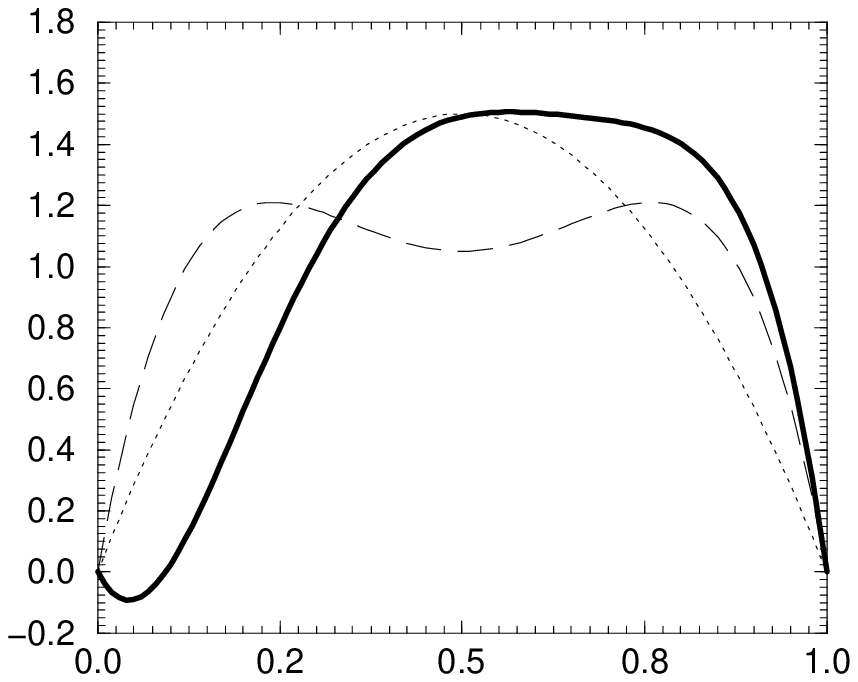}
\hspace*{1cm}
\includegraphics[width=6.5cm]{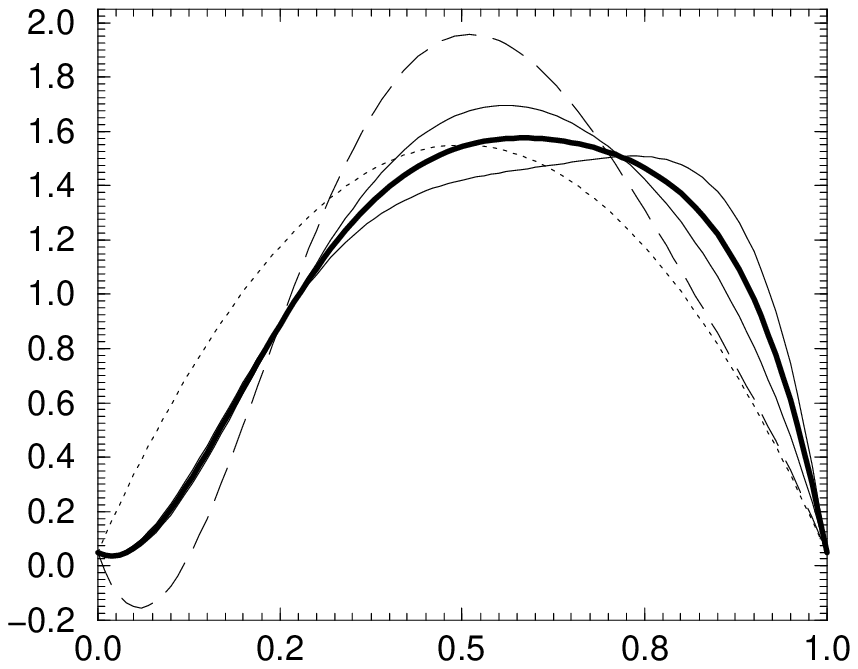}
\caption{\it Distribution amplitude $\phi_{\perp}(u)$ (bold curves) is plotted
for the case $\mu = 1\,GeV$ (left), and $\mu = \mu_b$ (right). In the right
figure,
the two other solid curves show the result of the
variation of the parameters $a_{1,2,3}^{\perp}(\mu_b)$ listed in App.B. Dashed
curve: in
the left figure, it shows this distribution amplitude for the case of $\rho$
meson as obtained in {\rm \cite{wfns}}; in the right figure, it is
$\phi_{\perp}(u)$
used in {\rm \cite{abs}}. Dotted curve in both figures shows the asymptotic
shape
$6u(1-u)$.}
\end{figure}

\noindent
Putting this into dispersion relations and taking the Borel transform of these
relations, we obtain the following expressions for our two form factors:
\bea
\label{tt1}
& &T_1(q^2) = \frac{m_b f_{K^*} m_{K^*} }{4 m_B^2 f_B}
\exp{\left[{\frac{m_B^2-m_b^2}{M^2}}\right]} \int_{0}^{1} \frac{du}{u}
\exp{\{\frac{\bar u}{u
M^2} (q^2 - m_b^2 - u m_{K^*}^2)\}} \times \nonumber\\
& &\left\{ \left[ u g_\perp^{(v)}(u,\mu) + \Phi_{\parallel}(u,\mu) +
\frac{1}{4} g_\perp^{(a)}(u,\mu) + \frac{m_b
f_{K^*}^{\perp}}{f_{K^*} m_{K^*}} \phi_{\perp}(u,\mu) \right] \Theta
[c(u,s_0)]\, +
\right. \nonumber \\
& &\left. \frac{1}{4} (q^2 + m_b^2 - u^2 m_{K^*}^2 )\, g_\perp^{(a)}(u,\mu)\,
\left[ \frac{1}{u M^2}\Theta[c(u,s_0)] + \delta[c(u,s_0)]\right] \right\}
\eea
\bea
\label{tt2}
& &T_2(q^2) = \frac{m_b f_{K^*} m_{K^*} }{4 m_B^2 f_B (m_B^2 - m_{K^*}^2)}
\exp{\left[\frac{m_B^2-m_b^2}{M^2}\right]} \int_{0}^{1} \frac{du}{u^2}
\exp{\{\frac{\bar u}{u
M^2}(q^2 - m_b^2 - u m_{K^*}^2)\}}\times \nonumber\\
& & \left\{ \left[ (m_b^2 - m_{K^*}^2 u^2 - q^2) \,( u g_\perp^{(v)}(u,\mu) +
\Phi_{\parallel}(u,\mu))\, +
\frac{m_b f_{K^*}^{\perp}}{f_{K^*} m_{K^*}} \phi_{\perp}(u,\mu)) + 2 u q^2
g_\perp^{(v)}(u,\mu)\right]\Theta[c(u,s_0)] + \right.  \nonumber \\
& &\left. \left[ \frac{1}{4}\, (m_b^2 - q^2 - u^2 m_{K^*}^2)^2 - u^2 m_{K^*}^2
q^2\right]  g_\perp^{(a)}(u,\mu)\, \left[\frac{1}{u M^2}\Theta[c(u,s_0)] +
\delta[c(u,s_0)] \right] \right\}
\eea
\noindent
where $M^2$ is the Borel parameter, $\bar u = 1 - u$ and $c(u,s_0) = u s_0 -
m_b^2 + q^2\bar u - u \bar u m_{K^*}^2$. Note that we used the
continuum subtraction prescription explained in \cite{bb}. Apart from that, the
above formulae agree with \cite{abs,aliev}. The
resulting form factors are shown in Fig 3.
\begin{figure}
\centering
\begin{minipage}[h]{\textwidth}
\centering
\subfigure[]{\includegraphics[width=11cm]{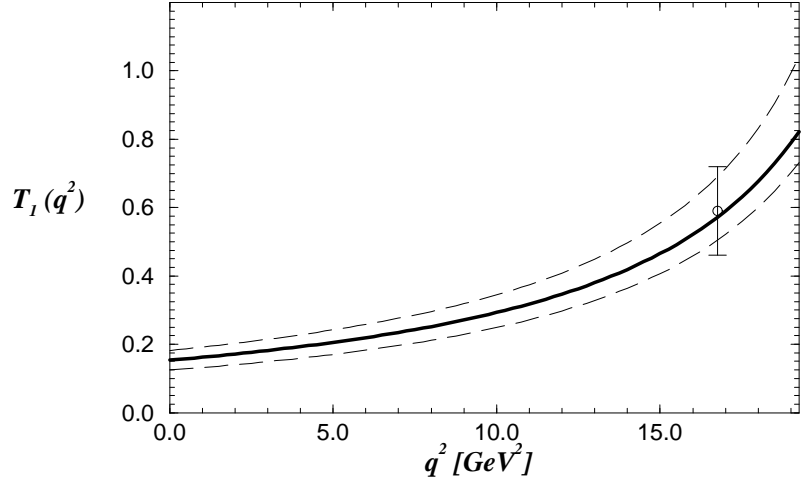}}
\end{minipage}
\centering
\begin{minipage}[h]{\textwidth}
\centering
\subfigure[]{\includegraphics[width=11cm]{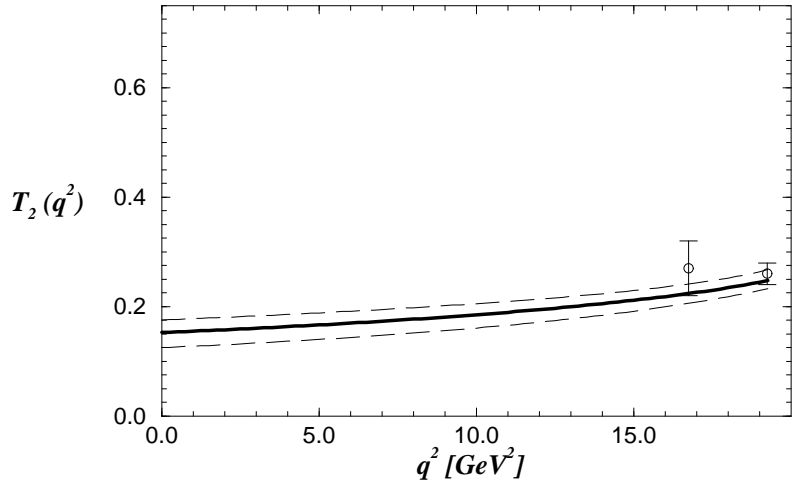}}
\end{minipage}
\caption{\it LC sum rule prediction for the $T_1(q^2)$ and $T_2(q^2)$ form
factor. Dashed lines represent the variation of the sum rule parameters as
indicated in appendix B.}
\end{figure}
\noindent

The results of the sum rule analysis are expected to be reliable for $q^2 -
m_b^2$ negative enough. For the case at hand, the range of reliability is below
$q^2 \sim 15~GeV^2$ \cite{abs}.
\newline
\noindent
{}From (4.11) and (4.12) (see also Fig.3), we extract the value of the form
factor:
\bea
T(0)=0.15 \pm 0.01 \pm 0.02 \pm 0.02^{th}
\eea
\noindent
where the first error comes from the variation of $m_b$ and the sum rule
parameters (Borel and the threshold parameters), the second error is related to
the variation of the wave
functions' coefficients (see App.B) and the third is `theoretical'
uncertainty of order $~ 15 \%$ (not included higher twist and radiative
corrections). To be on a safe side, we varied all
parameters that enter the calculation in their
largest reasonable ranges. This is particularly the case for $a_2^{\perp}(\mu)$
coefficient whose variation mainly contributes to the second error. Note that
on Fig.2 we displayed $\phi_{\perp}(u)$ for only one (`central') choice of
$a_2^{\perp}(1\, GeV)$. As it can be noticed, although with different shapes of
wave function, numerical consequences on the form factor $T(0)$, when compared
to \cite{abs}, are of rather minor importance (see Tab.1). This is due to the
fact that the value of
$f_{K^*}^{\perp}$ that we use is smaller than the one used in \cite{abs,aliev}
which compensates the
higher value of $a_2^{\perp}$ (see App.B). Let us also give some cross
results. Without SU(3) breaking, {\it i.e.} with the wave functions from
\cite{wfns} the
above sum rules yield:
\bea
T(0)=0.13 \pm 0.01 \pm 0.02 \pm 0.02^{th}
\eea
\noindent
We also evaluated the sum rule for $B\to \rho \gamma$\footnote{In \cite{cleo},
first candidates for $B\to \rho \gamma$ have been extracted.}
and the result is:
\bea
T(0)=0.12 \pm 0.01 \pm 0.01 \pm 0.02^{th}
\eea
\noindent
Of course the true result for this decay rate can be obtained only when
long distance effects from effective hamiltonian are included \cite{khodj}. In
the limit where they are
neglected, we have \cite{abs}:
\bea
{{\cal B}(B\to \rho \gamma) \over {\cal B}(\bk)} = \left| {V_{td}\over
V_{ts}}\right|^2 r^2(0) {m_b^2 + m_d^2 \over m_b^2 + m_s^2} \left(
{m_B^2 - m_{\rho}^{2} \over m_B^2 - m_{K^*}^2} \right)^3
\eea
\noindent
where for $r(0)= \vert T^{B\to \rho}(0) \vert / \vert T^{B\to K^*}(0)\vert$ we
obtain $r(0)\,=\,0.776 \pm
0.073$.

\section{Bounds with additional constraint}
\setcounter{equation}{0}

In the previous section we obtained the form factor at $q^2=0$, the value of
which is
estimated by including various uncertainties.
Now we can generate the bounds from sec. 3 with one lattice result at
$q^2 \sim q_{max}^2$ and LC sum rule result at $q^2 =0$, to see if something
more can be learned about the form factor behavior. These bounds are depicted
in Fig 4.

\begin{figure}
\vspace*{-.5cm}
\centering
\begin{minipage}[h]{\textwidth}
\centering
\subfigure[]{\includegraphics[width=11cm]{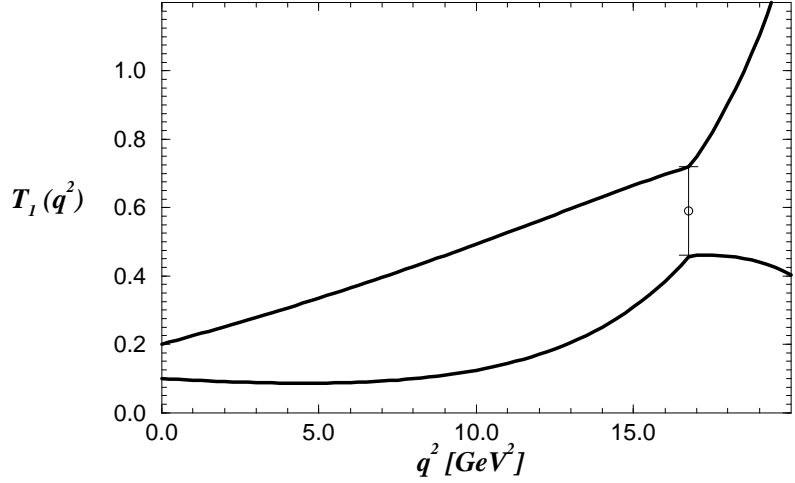}}
\end{minipage}
\vspace*{-.5cm}
\centering
\begin{minipage}[h]{\textwidth}
\centering
\subfigure[]{\includegraphics[width=11cm]{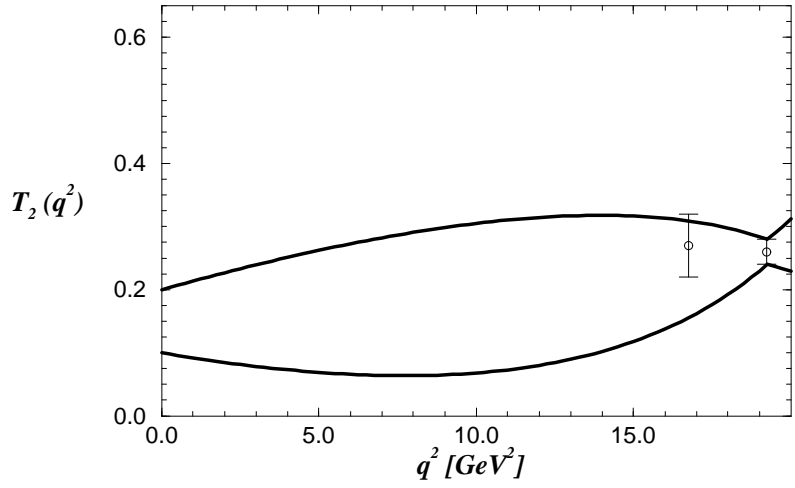}}
\end{minipage}
\caption{\it Unitarity bounds for the form factors $T_1(q^2)$ and $T_2(q^2)$
constrained by the lattice result at high $q^2$ and by the LC sum rule result
at
$q^2=0$.}
\end{figure}
\noindent
Now, the bounds are much better. It is obvious that for a clear conclusion on
the
form factors' behavior, we need some more points to constrain more the
intermediate region. For this, the lattice results are needed if we want to
treat both methods on equal footing. There are other sources of improvements of
this analysis: for the functions $\chi^{\pm}$, corresponding
two loop expressions can be calculated; try to seriously examine the effects of
subthreshold
singularities. At the same time, the bounds need to be relaxed for the value of
systematic uncertainties. Some of these uncertainties for the lattice results
are
included using the quadratic instead of the linear fit to extrapolate to $m_B$
\cite{ukqcd}.

However, even now from the two available results,
it is obvious that the fully flat behavior of $T_2$ can be discarded.
This does not contradict \cite{ape}, where the best fit was the pole one, with
the pole mass as a free parameter.
It is unlike that the nearest pole behavior can work in the full (large) region
for this decay. This would lead to the oversimplification of the contributions
of all possible singularities that lay behind the nearest
pole, especially in the large recoil region. We do not know strengths nor the
signs of these singularities,
but the least we can do is to have doubts on
their mutual cancellations and leave "$m_{pole}$" as a free parameter which is
to be fixed in fit with the data.
In the light of that remark, we note that the pole fit with $m_{pole} \sim
7~GeV$ reported in
\cite{ukqcd} is compatible with our bounds.
In fact the dipole/pole behavior is favored (of course not with
the nearest resonance masses). Specifically, in Fig1a. we can make the pole fit
with $T_1 (0) > 0.23$ and $m_{pole} \sim m_{B_s^*}$ while with the result from
Fig4a. this cannot be done and the double pole fit works. For instance, with
$T_{1,2} (0)=0.15$, the free parameter $m_{`dipole'}= (5.6 - 6.2)~GeV$ and
$m_{`pole'}= (6.4 - 7.0)~GeV$, where the variations of free parameters allow to
stay within lattice constrained bounds in the other end of the $q^2$ region.
This is illustrated in Fig.5.
That is what we obtain concerning the $q^2$ behavior of our form factors which
is important for the study of the $B\to K^*
\ell^+ \ell^-$ ({\it i.e.} for the off-shell photon).
\begin{figure}
\vspace*{-.5cm}
\centering
\begin{minipage}[b]{\textwidth}
\centering
\subfigure[]{\includegraphics[width=11cm]{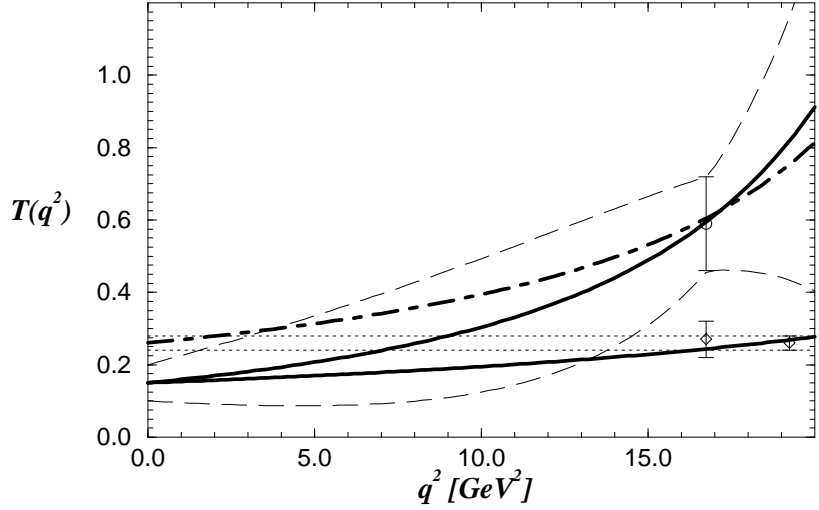}}
\end{minipage}
\caption{\it Form factor behavior: Bold solid lines show the double pole (pole)
behavior of $T_1(q^2)$ $(T_2(q^2))$ with
$T(0)=0.15$,
$m_{`dipole'}= 5.8\, GeV$, $m_{`pole'}= 6.7\, GeV$. Dot-dashed bold line shows
the nearest pole dominance for $T_1(q^2)$
compatible with {\rm (\ref{sclaw})} and $T_2(q^2)$ being flat (dotted lines).
Unitarity bounds for $T_1(q^2)$ are represented by dashed
lines.}
\end{figure}
\noindent
On the other hand, the extrapolation of the lattice results in mass at $q^2=0$
is surely a better way to extract the value of the form factor for $\bk$.
As we already mentioned such a scaling law is now available. Here,
we will not go through the lattice analysis again, but rather make an important
consistency check of the two approaches that we use in this paper.
First, one can calculate the heavy quark limit of the sum rules
\cite{abs}. We recalculated that too, and for very generous variations of the
sum
rule parameters we obtain:
\bea
T(0) m_H^{3/2} = (2.7 \pm 0.3)~GeV^{3/2}.
\eea
\noindent
Now we can make the fit with the available lattice data\footnote{where we again
use APE lattice results} as:
\bea
T(0) m_H^{3/2} = (2.7 \pm 0.3)~GeV^{3/2}~\left( 1 + {A\over m_H} + {B\over
m_{H}^2} \right)
\eea
for lattice attainable $m_H$, extrapolated to $q^2=0$ (with $m^{-3/2}$ scaling
law, see table2. in \cite{ape}) to fix A and B. The values of the coefficients
obtained in this way are:
\bea
A= - (2.1 \pm 0.2)~GeV ~~~ B= (1.0\pm 0.2)~GeV^2.
\eea
\noindent
These numbers are big, which indicates that the corrections to the infinite
quark mass limit are large.
Now, we simply extrapolate to the physical $B$ mass and the resulting value of
the form factor is:
\bea
T(0,m_B)=0.14\pm 0.02.
\eea
\noindent
If we include $\sim 10\%$ of errors to this result due to quenching, we obtain
the lower bound as $T(0,m_B)\geq 0.11$. So we can
reduce the uncertainty on $T(0)$ reported in the previous section ending up
with the still conservative result:
\bea
T(0)=0.15\pm 0.04.
\eea

Another consistency check of LC sum rules and lattice predictions can be made
for the strength of the nearest pole contribution, {\it i.e.} the residue $\sim
f^{\perp}_{B^*_s}~g_{B^*_s B K^*}$ which supposedly dominate the form
factor $T_1(q^2)$ behavior in the zero recoil region. We did not calculate
that but plan to do it with the new lattice data.

In the
table1., we enumerated some of the existing results insisting on the QCD
directly
based predictions. For completeness (comparison), some quark model predictions
are also given.

\begin{table}[H]
\begin{center}
\begin{tabular}{ccc}
Reference & $T(0)$ & $T_2(q^2_{max})$ \\ \hline
& & \\
This work &  $0.15\pm 0.04$ & $0.26\pm0.02$~\cite{ape}\\
& & \\
BHS \cite{bhs} & $0.10\pm 0.03$ & $0.33\pm 0.07$ \\
UKQCD \cite{ukqcd} & $0.15^{+.07}_{-.06}$ & $0.27^{+.02}_{-.01}$\\
APE \cite{ape} & $0.09\pm 0.02$ & $0.26\pm 0.02$  \\
LANL \cite{lanl} & $0.09\pm 0.01$ & $0.23\pm 0.01$  \\
& & \\
Colangelo {\it et al.}\cite{colangelo} & $0.19\pm 0.03$ & $0.14\pm 0.02$ \\
Aliev {\it et al.} \cite{aliev}& $0.15 $ & --  \\
Ball \cite{ball} & $0.19\pm 0.03$ & -- \\
ABS \cite{abs} & $0.16\pm 0.03$ & -- \\
Narison \cite{narison} & $0.15\pm 0.02$ & -- \\
DPR \cite{dpr} & $0.20$ & -- \\
& & \\
MNS \cite{melikhov} & $0.18\pm 0.05$ & $0.25-0.35$ \\
Stech \cite{stech} & $0.18$ & $0.35$ \\
Olsson, Veseli \cite{ov} & $0.13\pm 0.03$ & -- \\
\end{tabular}
\end{center}
\caption[]{The form factor $T(0)$ and $T_2(q^2_{max})$ obtained on the lattice
(quoted lattice results are obtained with pole/dipole scaling law), from LC
and traditional sum rules, and quark models. }
\end{table}

\section{Conclusion}
\setcounter{equation}{0}
In conclusion, let us briefly outline the result of our analysis. We generated
the unitarity bounds for the $\bk$ form factors accessible on the lattice. The
bounds constrained by lattice results only are not very restrictive. They will
become more useful with the new lattice data. Besides the heavy quark scaling
laws
in the small recoil region, the scaling law at maximum recoil is accounted for.
The former contradicts both form factors being pole dominated because of the
condition $T_1(0)=T_2(0)$. To have some
more insight in the form factors' behavior that can be drawn from this
analysis, we also recalculated the light cone sum rules for them. By including
the errors of theoretical uncertainties and varying the sum rule parameters in
their large reasonable ranges, we obtain that pole/double pole fit is in fact
better.
For the consistency check, we also evaluated the sum rule in the heavy quark
limit and fixed parameters for the extrapolation to the B-meson mass by
the existing lattice results. Such obtained results are very consistent with
the
results (bounds) obtained from LC sum rule. In generating
the bounds, the pole that exists for each form factor is properly accounted
for.
Then, the result that one of the form factors may be fitted by dipole may seem
a bit
surprising. This is not worrisome at all since the unconstrained bounds are
very wide (for instance $\vert T_1 (0)\vert < 10$), and the decisive
limitations come actually from the results taken to constrain these bounds. At
the end, the result that we extract from this analysis for the form factors at
$q^2 = 0$ and with all constraints included, falls in the conservative range
$0.15\pm 0.04$. With this, we can predict the hadronization ratio:
\bea
R_{K^*}&=&{{\cal B}(\bk)\over {\cal B}(b\to s  \gamma)} = 4 \left[ {m_B \over
m_b} {m_B^2 - m_{K^*}^2 \over m_B^2}\right]^3~\vert T(0)\vert^2 \nonumber \\
&=& (12.2 \pm 6.6)\%
\eea
\noindent
with the above formula being valid to leading order in QCD \cite{ciuc}. We also
give our prediction for the branching ratio for this decay, with $C_{7}=-0.333$
\cite{ali} and $\tau_B = 1.6~ps$ \cite{pdg}:
\bea
{\cal B} (\bk)&=& \vert V_{ts}^* V_{tb}\vert^2~(2.41\pm 1.27)\times 10^{-2}
\nonumber \\
&=& (4.5 \pm 3.2) \times 10^{-5}
\eea
\noindent
$\vert V_{ts}^* V_{tb}\vert$ is also taken from \cite{pdg}. With these
parameters, we can extract the experimental values of the form factor: $T(0)=
0.17\pm 0.06$. New theoretical and computing improvements are thus more than
needed.
\vskip1cm
{\bf Acknowledgement:}
\vskip1cm
It is a pleasure to thank J.-P. Leroy for motivating discussions, O. P\`ene for
his
important comments. I am particularly indebted to V. M. Braun for pointing out
to me the importance of their findings in \cite{wfns} and very
helpful remarks. I also thank T. Bhattacharya and D. Melikhov for informations
on their respective results. Finally, I would like to thank G. Martinelli
and the University of Rome "La Sapienza", where a part of this work was done.

\newpage

\appendix
\renewcommand{\theequation}{\Alph{section}.\arabic{equation}}
\section*{Appendices}

\section{Explicit derivation of the unitarity bounds}
\setcounter{equation}{0}

In this appendix we outline the way to get the unitarity bounds on the form
factors with their values known at several points as constraints. The `pure'
unitarity bounds can be obtained as usual (see any of references
\cite{bds1,bds11,caprini,bds2,oku,lellouch,damir}):
\bea
\label{pure}
\vert T(z)\vert \leq {1\over \vert \phi(z)\vert}{1\over \sqrt{1-z^2}}
\eea
\noindent
and they are very weak for the case we study. Suppose now that form factor is
known at several points $\{z_i\}_{1}^{n-1}$. Recall that the functions we
consider are real and analytic
in $\vert z\vert = 1$ belonging to unit sphere of ${\cal H}^2_w$ {\it i.e.}:
\bea
\label{const}
\parallel T w\parallel_{L^2} \leq 1
\eea
\noindent
For example, when the form factor is known at $z=\{0,z_0\}$, we look for the
analytic function
which takes the two prescribed values and
satisfies the constraint \ref{const}, which is equivalent to methods used in
\cite{bds1,oku}. For
this to
work, the sequence of points $z_i$ should be uniformly separated
\cite{duren}. If this is not fully
satisfied, the probabilistic approach can be adopted as it was done in
Ref.\cite{lellouch}.

The most general functional form which satisfies above mentioned constraints
is:
\bea
T(z)=\sum_{i=1}^{n} T(z_i) {B_i(z)\over B_i(z_i)}~ +~ B(z) g(z)
\eea
\noindent
which can be simply checked employing the usual definitions:
\bea
B(z) = \prod_{i=1}^{n} \frac{z-z_i}{1-z z_i} ~~~ B_i(z) =
\prod_{i\neq j}^{n} \frac{z-z_j}{1-z z_j}
\eea
\noindent
Obviously, the first term in $T(z)$ satisfies both constraints
({\it i.e.} prescribed values and (\ref{const}) ) while the second term
contains an unknown function $g(z)$. The aim is then to look for the $g(z)$
belonging to $H_w^2$,
closest to the given kernel {\it i.e.} \cite{guiasu,caprini}:
\bea
\mu^2 = (\min\limits_{g\in H_w^2} \Vert g - \psi \Vert )^2
\eea
where the kernel:
\bea
\psi =  \sum_{i=1}^{n} {T(z_i) \over B_i(z_i)} {1 - z z_i \over z - z_i}
\eea
\noindent
is just the first term of $T(z)$ divided by $B(z)$.
It is convenient to solve this problem by considering its dual \cite{duren}:
\bea
\mu^2 = \sup\limits_{G\in H^2,\, \Vert G \Vert \leq 1} \vert {1 \over 2 \pi }
\int_{0}^{2\pi} d\theta~G(\theta)\psi(\theta)\phi(\theta) \vert
\eea
\noindent
$G(z)=\sum G_k z^k$ is surely inside the unit sphere of $H^2$ if $\sum G_k^2
\leq 1$ which again reflects the constraint (\ref{const}). Then we simply apply
the residua theorem to calculate integrals and
obtain:
\bea
\label{capr}
\mu^2 = \sup\limits_{\sum G_k^2 \leq 1} \sum^{\infty}_{k=0} G_k \xi^{(n)}_{k}
\eea
where by $\xi{(n)}_k$ we compactly wrote the result of integration of each term
of the power series, and $n$ stands for the general case
of $n-1$ known form factor values at fixed $z=z_n$ which we then vary to obtain
the functional dependence that we are interested in. Now, using the Schwartz
inequality, we eventually obtain:
\bea
\label{imp}
\sum_{k=0}^{\infty}\left[ \xi^{(n)~2}_{k}\right]^{1/2} \leq 1
\eea
\noindent
It is very easy to show that for $n=1$ we obtain the pure unitarity bounds
(\ref{pure}).

For example, when $n=3$ {\it i.e.} the form factor value is known at
$z_{1,2}=\{0,z_0\}$ and we look for its functional dependence at $z_3=z$, in
(\ref{capr}) we have:
\bea
\left({T(0)\over z_0 z} + T(z_0) {1 - z_0 z\over z_0 - z}{1 - z_0^2 \over z_0}
+ T(z){1 - z_0 z\over z - z_0}{1 - z^2 \over z} \right)G_0  \nonumber \\
 +\sum_{k=1}^{\infty} \left( {T(z_0)\over z_0-z}(1-z_0 z) (1 - z_0^2) z_0^{k-1}
+ {T(z)\over z-z_0}
(1-z_0 z) (1 - z^2) z^{k-1} \right) G_k
\eea
\noindent
and by inserting it to (\ref{imp}), we obtain the bounds given in
Eq.(\ref{bds3}). The case $n=2$ that appears in the paper, is obtained in very
much the same way.

Note that each time the form factor is multiplied by a proper Blaschke factor
to account for the pole. In order not to make it heavier then necessary, this
is implicit in all the formulae displayed so far. To have some feeling about
the numerical analysis, we give some significant numbers:
\begin{itemize}
\item{For the form factor $T_1(q^2)$, the measured lattice point is at
$q^2=16.8~GeV^2 \to z=0$ when $N=1.130$ and the pole position in z-plane is at
$q^2=29.38~GeV^2 \to z_{pole1}=-0.2188$. Also $z(q^2=0)=0.1443$ was used.}
\item{For the form factor $T_2(q^2)$, the measured lattice point is at
$q^2_{max}= 19.2~GeV^2 \to z=0$ with $N=1$; the pole position $q^2=32.15~GeV^2
\to z_{pole2}=-0.2811$ and $z(q^2=0)=0.1742$.}
\end{itemize}

\newpage

\section{Distribution amplitudes and numerical parameters used in LC sum rules}
\setcounter{equation}{0}

For completeness, we display the wave functions that we use in the
evaluation of the sum rules. We show only the steps which are particularly
important for our analysis. The rest can be found in {\it e.g.} \cite{wfns}. To
leading twist accuracy, there are two functions that are independent:

\bea
\phi_{\perp,\parallel} (u,\mu) &=& 6 u (1-u)\left[ 1 +
a_1^{\perp,\parallel}(\mu) \zeta + a_2^{\perp,\parallel}(\mu) \left(
\zeta^2 - \frac{1}{5}\right) + a_3^{\perp,\parallel} (\mu) \left( \frac{7}{3}
\zeta^3 - \zeta
\right) \right].
\eea
\noindent
To the same accuracy, the other functions that figure in Eq.(4.6,4.7,4.8), {\it
i.e.}
$g_{\perp}^{(v,a)},\Phi_{\parallel}$ are not independent. They are expressed in
terms of $\phi_{\parallel}(u,\mu)$ (see \cite{wfns,abs}). Their explicit forms
read as follows:
\bea
g_{\perp}^{(v)} (u,\mu) &=& \frac{3}{4} \left( 1 + \zeta^2 \right) +
a_1^{\parallel}(\mu) {\zeta^3 \over 3}  +
a_2^{\parallel}(\mu) {1 \over 40} \left( 15 \zeta^4 - 6 \zeta^2 - 1 \right) +
a_3^{\parallel}(\mu) {\zeta^3 \over 2} \left( {7
\over 5} \zeta^2 - 1\right) ,\nonumber \\
g_{\perp}^{(a)} (u,\mu) &=& 6 u (1-u) \left[ 1 + a_1^{\parallel}(\mu) {\zeta
\over 3} + a_2^{\parallel}(\mu) {1 \over 6}
\left( \zeta^2 - {1 \over 5} \right) + a_3^{\parallel}(\mu) {1 \over 10}\left(
{7 \over 3} \zeta^3 - \zeta\right)
\right],\nonumber \\
\Phi_{\parallel} (u,\mu) &=& {3 \over 2} u ( 1 - u) \left[ \zeta +
a_1^{\parallel}(\mu) {1 \over 3} \left(2 \zeta^2 -
1\right) + a_2^{\parallel}(\mu) {\zeta \over 2} \left( \zeta^2 - {11\over
15}\right) + a_3^{\parallel}(\mu) {1\over 10}
\left( {28\over 3}\zeta^4 - 9 \zeta^2 + 1\right)\right] \nonumber \\
& &
\eea
\noindent
where $\zeta = 2 u - 1$. In the case of the $\rho$ meson $a_1=a_3=0$, and
$a_2^{\perp}(1\,GeV)= 1.50\pm 0.75$, $a_2^{\parallel}(1\,GeV)=
1.35\pm 0.75$ \cite{wfns}. The coefficients $a_{1,3}$ give rise to
antisymmetric parts of the wave functions which acquire nonzero
values upon SU(3) breaking. To obtain the values of the coefficients, we use
the results of \cite{zhitn} for corresponding moments $\lan
\zeta^n \ran$:
\bea
{\lan \zeta^2 \ran^{\perp}_{K^*} \over \lan \zeta^2 \ran^{\perp}_{\rho}} = 0.75
, ~~~
{\lan \zeta^2 \ran^{\parallel}_{K^*} \over \lan \zeta^2
\ran^{\parallel}_{\rho}} = 0.85,
\eea
\noindent
for $a_2^{\perp,\parallel}(1\,GeV)$, and
\bea
{\lan \zeta^3 \ran^{\perp,\parallel} \over \lan \zeta \ran^{\perp,\parallel}} =
0.50 -0.55 ,~~~ \lan \zeta \ran^{\perp,\parallel} = 0.15
- 0.20,
\eea
\noindent
for $a_{1,3}^{\perp,\parallel}(1\,GeV)$. Thus obtained values are then evoluted
to $\mu_b^2 \simeq 5\,GeV$, so that we obtain the following
results (ranges of values) for the coefficients:
\bea
a_1^{\parallel} (\mu_b) = 0.64 \pm 0.09, ~~a_2^{\parallel} (\mu_b) = 0.31 \pm
0.37, ~~a_3^{\parallel} (\mu_b) = 0.38 \pm 0.15;
\nonumber \\
a_1^{\perp} (\mu_b) = 0.69 \pm 0.10, ~~a_2^{\perp} (\mu_b) = 0.21 \pm 0.41,
{}~~a_3^{\perp} (\mu_b) = 0.42 \pm 0.16.
\eea
The values of these coefficients are varied in their respective ranges and the
form factor results are relatively insensitive to it, as we reported in (4.13).
This is true provided SU(3) breaking is carried also for the constants $f_V$
and especially $f_V^{\perp}$. Again, from \cite{wfns}: $f_{\rho}^{exp} =
(205 \pm 10)\,MeV$, $f_{\rho}^{\perp}= (160 \pm 10)\, MeV$, and with
$(f_{K^*}^{(\perp)} / f_{\rho}^{(\perp)})^2 =1.1$ from \cite{zhitn},
we have:
\bea
f_{K^*} = (215 \pm 10)\, MeV ~~~ f_{K^*}^{\perp} = (170 \pm 10)\, MeV
\eea
\noindent
So far the coefficients $a_i$ and $f_V^{\perp}$ ($V$ stands for a vector meson)
were obtained in the QCD sum rule
approach. There are very few lattice calculations which were performed using
small lattices
and a relatively poor statistics. For the case of pion distribution amplitude,
the authors of
Ref.\cite{marsac} studied the second moment for which we can extract the value
of $a_2^{\pi}(1\,GeV)=0.44\pm 0.53$. The errors are very large and they are
mostly due to the
modest statistics. Only recently this result was improved in the study of
\cite{desy}\footnote{In this reference
the authors also calculated the moments for the $\rho$ meson. The lattice
estimate of $f_V^{\perp}$ and the comparison
of these results with the results of QCD sum rules
will be addressed elsewhere.} from which we
extract $a_2^{\pi}(1\,GeV)=0.40\pm 0.27$. It is rather difficult to compare
these results to that obtained
from the QCD sum rules $a_2^{\pi}(1\,GeV)=0.43$, although they agree well at
first sight.
The sum rule study \cite{zhitn} of the leading twist distribution amplitude for
the pion
needs also to be updated as it was done for the case of the rho-meson
\cite{wfns}.

In numerical analysis of the sum rules, the following values are used:
\begin{itemize}
\item{$b$ quark mass varied - $m_b=(4.7 \pm 0.1)\, GeV$}
\item{threshold parameter varied - $s_0 = (34\pm 1)\, GeV$}
\item{Borel parameter $M^2 = (6 - 10)\, GeV^2$}
\end{itemize}
\noindent
For the sum rule to estimate the value of $f_B$, also $\lan \bar q q\ran
(1~GeV) = (-245\pm 10 MeV)^3$, and $\lan \bar q g \sigma G q \ran (1~GeV) =
0.65~GeV^2~\lan \bar q q\ran (1~GeV)$ was used (the values taken from
\cite{bb}). The sum rule for $f_B$, is taken from \cite{abs} where the
radiative corrections are not included. This hopefully cancels some
uncertainties since these corrections (though needed) are not calculated for
the form factors yet\footnote{While this paper was in writting, the authors of
Ref.\cite{stefan}
calculated the radiative corrections to
the LC sum rule for the $f^+(q^2)$ form factor which enters $B\to \pi \ell
\nu$. Indeed,
their results show that the uncertainties due to these corrections
almost exactly cancel against the corresponding corrections for $f_B$.}.
The stability of the sum rules when the Borel parameter is varied is shown in
Fig.6. Also the change due to the subtraction \cite{bb}, with and without
delta-function term is shown on this figure.

\newpage

\begin{figure}
\centering
\includegraphics[width=7.5cm]{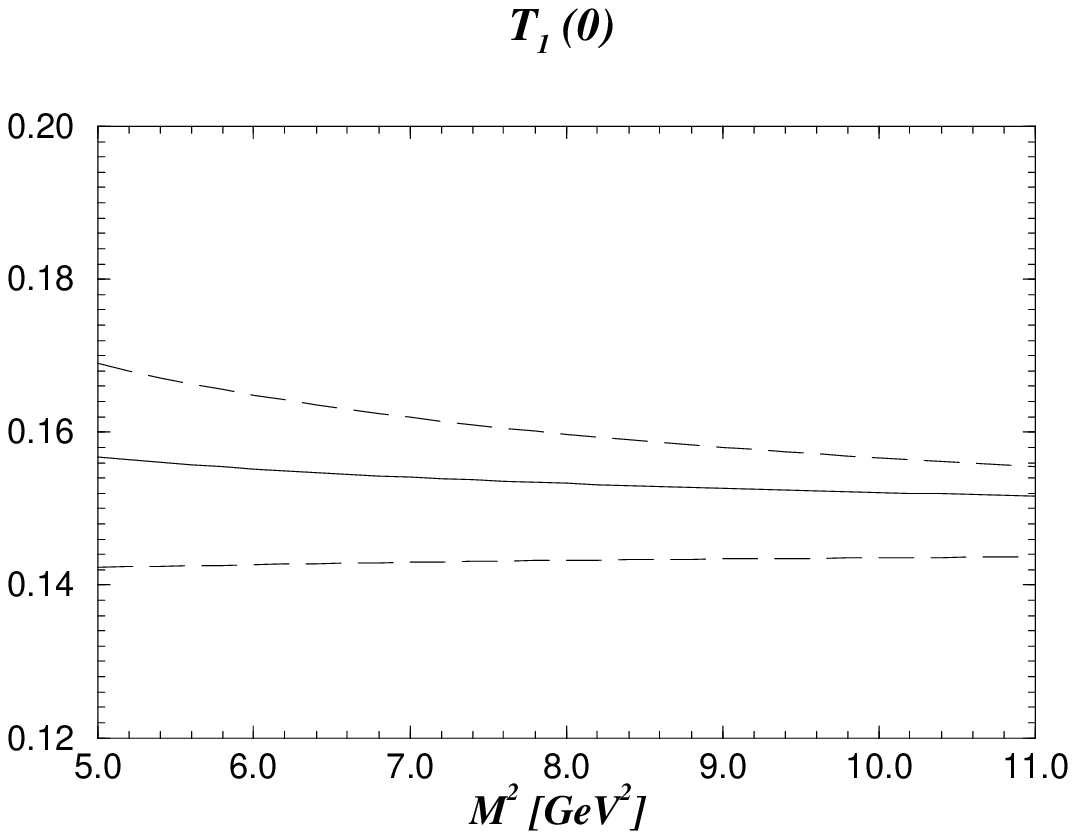}
\includegraphics[width=7.5cm]{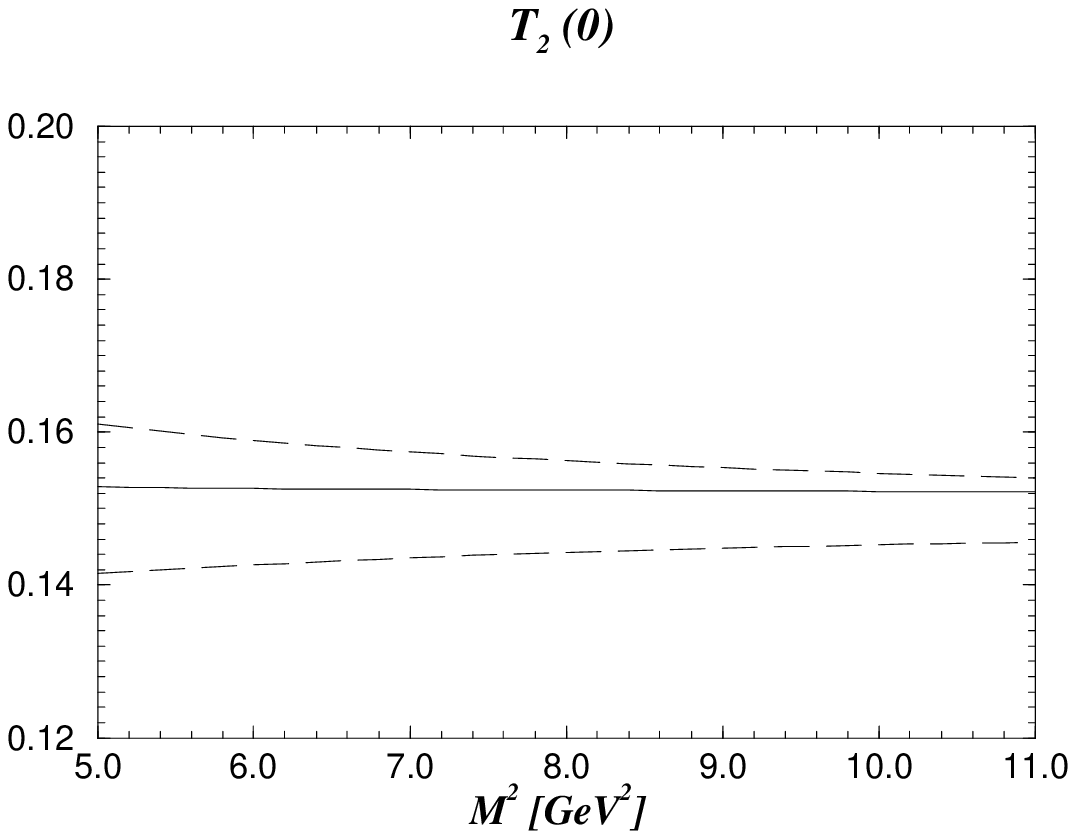}
\end{figure}
\begin{figure}
\centering
\includegraphics[width=7.5cm]{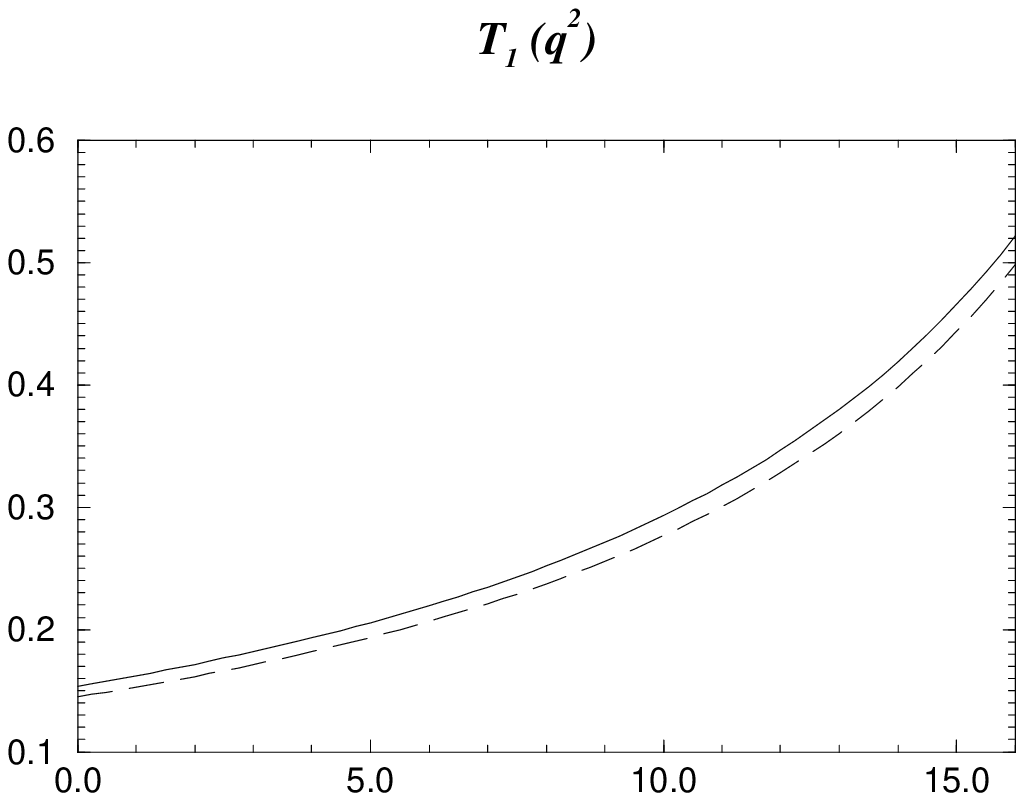}
\includegraphics[width=7.5cm]{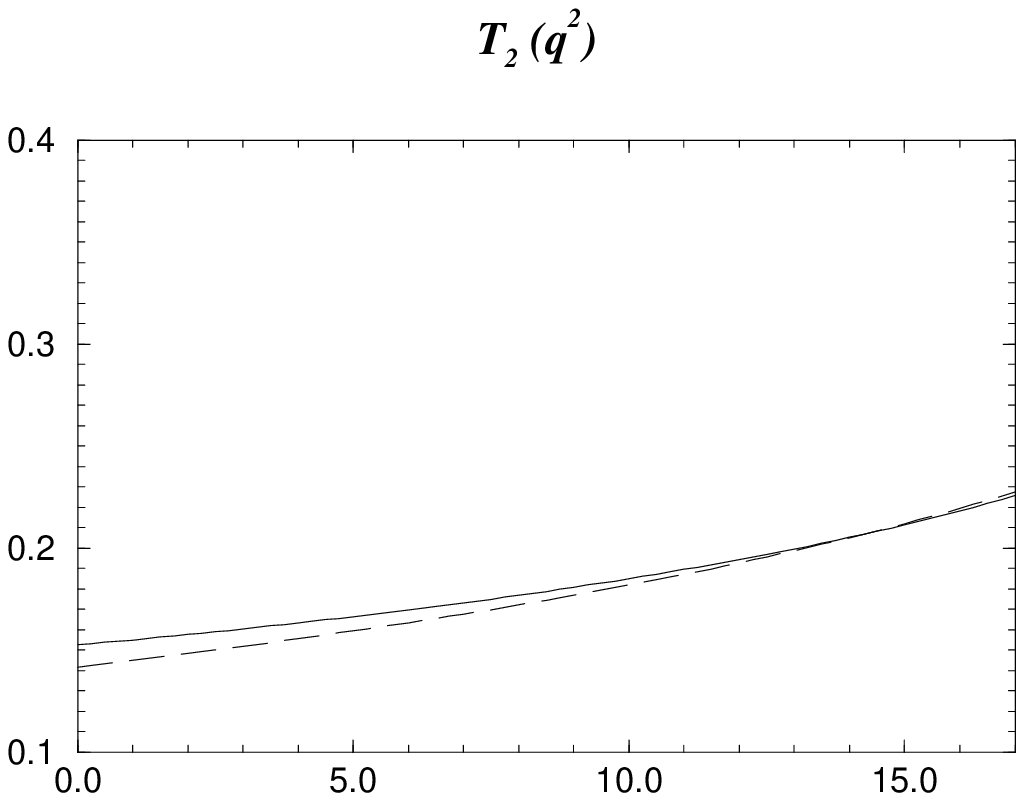}
\caption{\it In the figures a) and b) the form factors' dependence on the Borel
parameter is plotted. Dashed lines show the variations of $m_b$ and $s_0$. In
the figures c) and d) we plot the form factor behavior for $T_{1}(q^2)$ and
$T_2(q^2)$ for $M^2=8~GeV^2$ and $m_b=4.8~GeV$ with and without (dashed) delta
function term for the continuum subtraction in LC sum rules.}
\end{figure}

\newpage


\end{document}